\documentclass[pre,twocolumn,notitlepage,superscriptaddress,color,showpacs]{revtex4-1}
\usepackage[pdftex]{graphicx}
\usepackage{amsmath,amsfonts,amssymb,verbatim,color,wasysym,cleveref}

\begin{document}
\title{Threshold cascade dynamics on coevolving networks}
\author{Byungjoon Min}
\email{bmin@cbnu.ac.kr}
\affiliation{Department of Physics, Chungbuk National University, Cheongju, Chungbuk 28644, Korea}
\affiliation{Research Institute for Nanoscale Science and Technology, 
	Chungbuk National University, Cheongju, Chungbuk 28644, Korea}.
\author{Maxi San Miguel}
\email{maxi@ifisc.uib-csic.es}
\affiliation{IFISC, Instituto de Física Interdisciplinar y Sistemas Complejos (CSIC-UIB), 
	Campus Universitat Illes Balears, E-07122, Palma, Spain}	

\date{\today}

\begin{abstract}
We study the coevolutionary dynamics of network topology and social complex contagion using 
a threshold cascade model. Our coevolving threshold model incorporates two mechanisms: 
the threshold mechanism for the spreading of a minority state such as a new opinion, idea, or innovation
and the network plasticity, implemented as the rewiring of links to cut the connections between 
nodes in different states. Using numerical simulations and a mean-field theoretical analysis, 
we demonstrate that the coevolutionary dynamics can significantly  affect the cascade dynamics. 
The domain of parameters, i.e., the threshold and  mean degree, for which global cascades occur 
shrinks with an increasing network plasticity, indicating that the rewiring process suppresses 
the onset of global cascades. 
We also found that during evolution, non-adopting nodes form denser connections, resulting 
in a wider degree distribution and a non-monotonous dependence of cascades sizes
on plasticity.
\end{abstract}
\maketitle

\section{Introduction}

Understanding collective  interactions among agents is crucial for predicting the behavior  
of complex systems~\cite{granovetter,watts2002,centola,centola2018,battiston2021}. Recently, studies  
of group and higher-order interactions have received significant interest in the study of the  statistical physics 
of complex systems~\cite{laurent2020,centola2,battiston2021,levine2017}. 
Social contagion is one of the most interesting examples of group interactions, underlying the spread 
of information, fads, opinions, or behaviors~\cite{schelling,granovetter,centola,watts2002,gleeson2007,karsai2014,yy2014,auer,kook2021}.
Unlike the simple contagion process for the spread of infectious diseases which occurs via pairwise 
interactions~\cite{pastor}, social complex contagion~\cite{centola,centola2,centola2018} usually 
requires simultaneous interactions with multiple neighbors. 
The threshold model  is a pioneering work in the field of complex contagion describing cascading 
dynamics~\cite{schelling,watts2002,granovetter,gleeson2007}. It is a binary-state model in which 
the adoption  of an initial minority state by a node in an interaction network requires  that 
the fraction of neighboring nodes that have already adopted that state exceeds  a threshold value.
Cascade phenomena described by this model can represent not only the spread of social behaviors but 
also the transmission of neural signals~\cite{friedman}, error propagation in financial 
markets~\cite{kmlee}, and the collapse of power grids~\cite{motter2002}.

Although many studies have been conducted on threshold cascade
models~\mbox{\cite{watts2002,gleeson2007,hackett2011,lee2014,bmin2018_dual,abella2022,lee2023}}, including competition of 
simple and complex contagion processes~\cite{bmin2018,Czaplicka2016,diaz}, most have focused only on 
the dynamics on static networks~\cite{watts2002,gleeson2007,hackett2011,lee2014,bmin2018_dual,abella2022,lee2023}.
However, real-world complex systems change their connection patterns and the network of interactions 
changes dynamically~\cite{holmebook,adaptive_book,gross2008,funk2009}. 
In this respect, some studies have attempted to analyze coevolutionary dynamics, that is, dynamical 
processes in which the time evolution of the states of the nodes and the evolution of the network 
topology are dynamically coupled. These include coevolving voter 
models~\cite{holme2006,vazquez2008,sudo2013,diakonova2014,diakonova2015,bmin2016,raducha,bmin2019}, 
coevolving spin systems~\cite{mandra2009,raducha2018}, coevolving models of opinion formation~\cite{fu,su},
epidemic models of adaptive networks~\cite{gross2006,marceau2010,shaw2008},
coevolving models of cultural evolution~\cite{coevAxelrod1,coevAxelrod2},
and game theoretical models~\cite{CG}. 
While we here focus on the coevolution of node states and network topology,
there have been studies that address the coevolution between different dynamical processes
in a static network \cite{coelho,granell,pires,fang}.
In cases where cascading dynamics are coupled with the evolution of the network structure, it is 
essential to understand the coevolutionary dynamics of the network topology and threshold dynamics. 
However, only a few studies have been conducted on  models of coevolutionary dynamics including group 
or collective interactions~\cite{mandra2009,lambiotte2011}. Here, we attempt to understand the behavior 
of the threshold cascade model by incorporating the adaptive dynamics of the network topology. This is a 
tool for a better understanding of the comparison of threshold models with empirical 
data~\cite{centola,karimi-2013,karsai2014,rosenthal-2015,karsai-2016,mnsted-2017,unicomb-2018,guilbeault-2021,aral}.

In this study, we propose a coevolving threshold cascade model, where the nodes are in two possible 
states and can redefine  their connections in the network depending on the dynamical states of the nodes.
Initially, only a few seed nodes in a network are in a minority state that can represent new 
information, opinions, or innovations that might spread into the system.  According to the threshold 
process, given a node $i$ in the initial majority state, if the fraction of its neighbors that are 
already in the new initial minority state exceeds a certain threshold $\theta$, the node $i$ changes 
state,  and becomes ``adopting''. In addition, by following the homophilic tendencies observed in 
society~\cite{homo1,homo2}, an agent may reduce its social ties with individuals who are in an opposite 
state and establish new connections at random with agents who share the same state. To be specific, when 
a node $i$ is adopting, then a non-adopting node from the neighbors of node $i$ breaks its link with 
node $i$ and establishes a new link with a non-adopting node in the network. Therefore, the evolution 
of the network topology by link rewiring is coupled with the complex contagion processes so that 
the network structure constantly evolves in response to changes in the behavior of its constituents.
The main result obtained from simulations of this model, which is well described by an appropriate 
mean-field theoretical approach, is that the rewiring process can suppress the emergence of global 
cascades by a mechanism of the segregation of adopting~nodes. 

\section{Model}

\begin{figure}
\includegraphics[width=\linewidth]{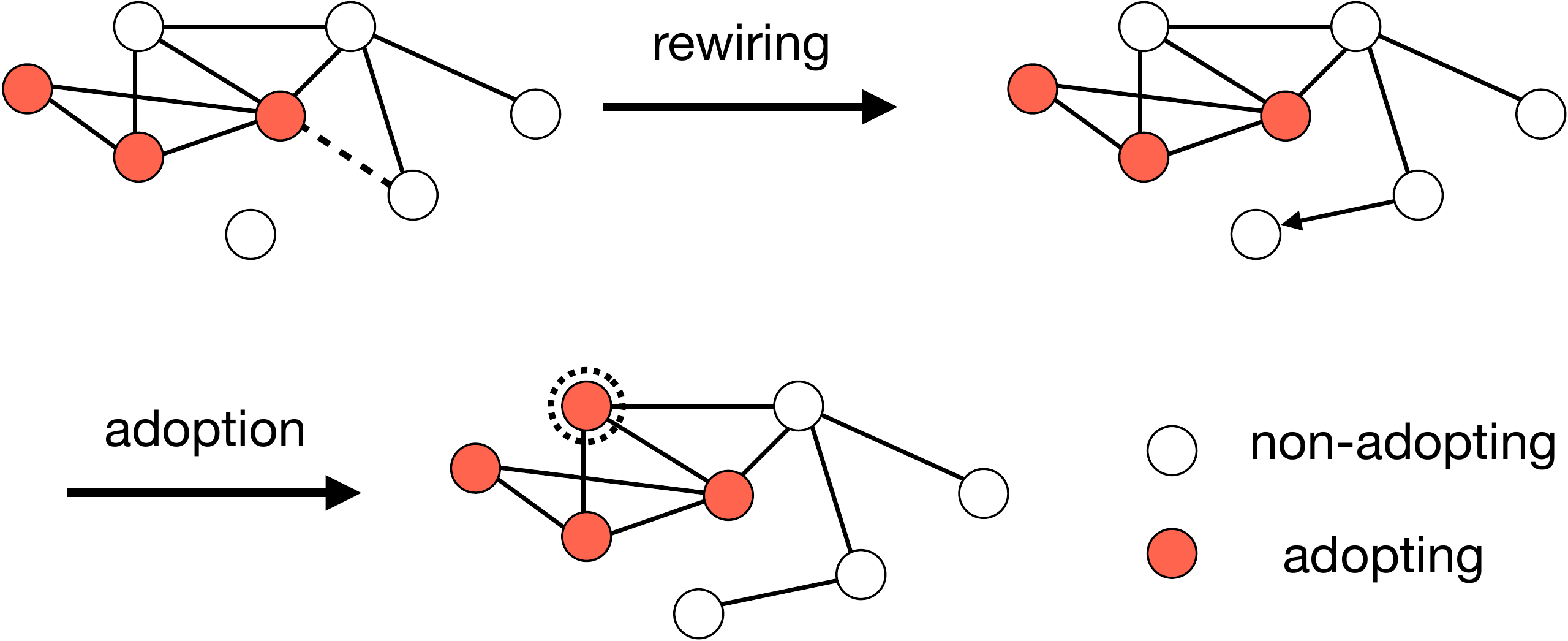}
\caption{An  
 example of the evolution rules of the coevolutionary dynamics of a threshold cascade model. 
A connected pair of an adopting (filled circles) and a non-adopting 
node (open circles) is removed with a probability $p$, and the non-adopting node establishes 
a connection to a new node that is not adopting, chosen randomly from the entire network. 
In addition, a non-adopting node becomes adopting if the fraction of adopting neighbors 
is larger than the threshold $\theta$. Once a node becomes adopting, the node is then permanently in this state. 
}\label{figure1}
\end{figure}

We consider a coevolving threshold cascade model as shown in Figure~\ref{figure1}. The coevolutionary threshold 
model consists of two parts: (i) the rewiring of links and (ii) the adoption of a new state 
(opinion, idea, or innovation). Dynamics start from seed node initiators: a small fraction 
$R_0$ of randomly selected adopting nodes. Furthermore, the dynamics proceed by the specific rules below.

Link rewiring  adaptively changes the structure of the network representing the situation 
in which an agent meets a new possible state, but does not want to adopt it.  At each time step 
of a random sequential update, each link that connects a pair of an adopting and a non-adopting 
node is removed with probability $p$. In addition, the non-adopting node that loses a link establishes 
a new link with a node that is not  currently adopting,  chosen randomly from the entire network. 
The parameter $p$, called the network plasticity, is a measure of the ratio of the timescales of 
network evolution to the adoption dynamics.

The adoption of a new state is a complex contagion process following the dynamics of Granovetter's and 
Watts' threshold model \cite{granovetter,watts2002}, where a non-adopting node becomes adopting if the 
fraction of its adopting neighbors exceeds a threshold $\theta$. We assume that each node has the same 
threshold $\theta$. Once nodes are adopting, their adopting state remains permanently. 
The two processes of link rewiring and adoption proceed until there are no active links connecting 
a pair of adopting and non-adopting nodes in a network.

\section{Results}

\subsection{On a Static Network}

To establish a benchmark for comparison, we begin by analyzing the threshold dynamics 
on a static (non-adaptive) network. This is a well-established model to explain the 
onset of the extensive size of the cascade of adoption from a few seed nodes, referred 
to as a ``global cascade'' \cite{watts2002,gleeson2007}. Typically, the global cascade occurs 
in a specific domain of two parameters: network connectivity and threshold. 
For instance, in Erd\"os--R\'enyi (ER) graphs, when the average degree $z$ is less than 
the percolation threshold $z_1$ of the graphs, global cascades do not occur, as there 
is no giant connected component. In addition, when $z$ is greater than a second 
threshold $z_2$ which depends on the threshold $\theta$, the nodes that exceed 
their threshold are  rare because the network becomes too dense. 
Therefore, global cascades can occur only in the range between $z_1$ and $z_2$.

For local tree-like networks, the transition lines in the parameter space between the global 
cascade and no cascade domains can be precisely identified using a mean-field 
analysis \cite{watts2002,gleeson2007}. On a random graph, the average fraction of the adopting nodes in a
stationary state, called the cascade size $R$, is given by the probability 
of a randomly selected node to become adopting. The size $R$ can be obtained by
approximating the network as a tree, with a chosen node as the root and 
considering the cascade of adoption towards the root. For a fixed degree distribution
$P(k)$ and initial seed fraction $R_0$, such a probability is given by \cite{gleeson2007}:
\begin{align}
R&= R_0 + (1-R_0)  \\
&\times	\sum_{k=0}^{\infty} P(k)  \sum_{m=0}^{k} \binom{k}{m} q_{\infty}^m (1-q_{\infty})^{k-m} F(m/k,\theta), \nonumber
\end{align}
where $q_{\infty}$ represents the probability that a node, reached via a randomly selected link,
is adopting in the stationary state and $F(m/k,\theta)$ is the threshold function.
To be specific, if $m/k > \theta$, $F(m/k,\theta)=1$, otherwise  $F(m/k,\theta)=0$.

The probability $q_{\infty}$ is computed by solving the following self-consistency 
equation iteratively \cite{gleeson2007},
\begin{align}
\label{eq:qn}
q_n &= q_0 + (1-q_0) \sum_{k=0}^{\infty} \frac{k P(k)}{z}  \\
& \times \sum_{m=0}^{k-1} \binom{k-1}{m} q_{n-1}^m (1-q_{n-1})^{k-m-1} F(m/k, \theta), \nonumber
\end{align}
where $q_n$ is the probability of step $n$ and $q_0=R_0$. In the limit $n \rightarrow \infty$,
we can obtain the probability $q_\infty$ in the steady state. 
In addition,  mean-field theory predicts the necessary conditions for global cascades
from the linear stability analysis of a trivial fixed point $q_{\infty}=0$ 
in the limit $R_0 \rightarrow 0$ as:
\begin{align}
\label{pc}
\sum_{k=1}^{\infty} \frac{ k (k-1)}{z} P(k) F(1/k,\theta) > 1.
\end{align}
Using this transition point and the size of adopting nodes predicted from the above theory  
as benchmarks, we will now analyze how they are modified by the coevolutionary adaptive dynamics of the network.

\subsection{Segregation of Adopting Nodes via Link Rewiring}

\begin{figure}
\includegraphics[width=\linewidth]{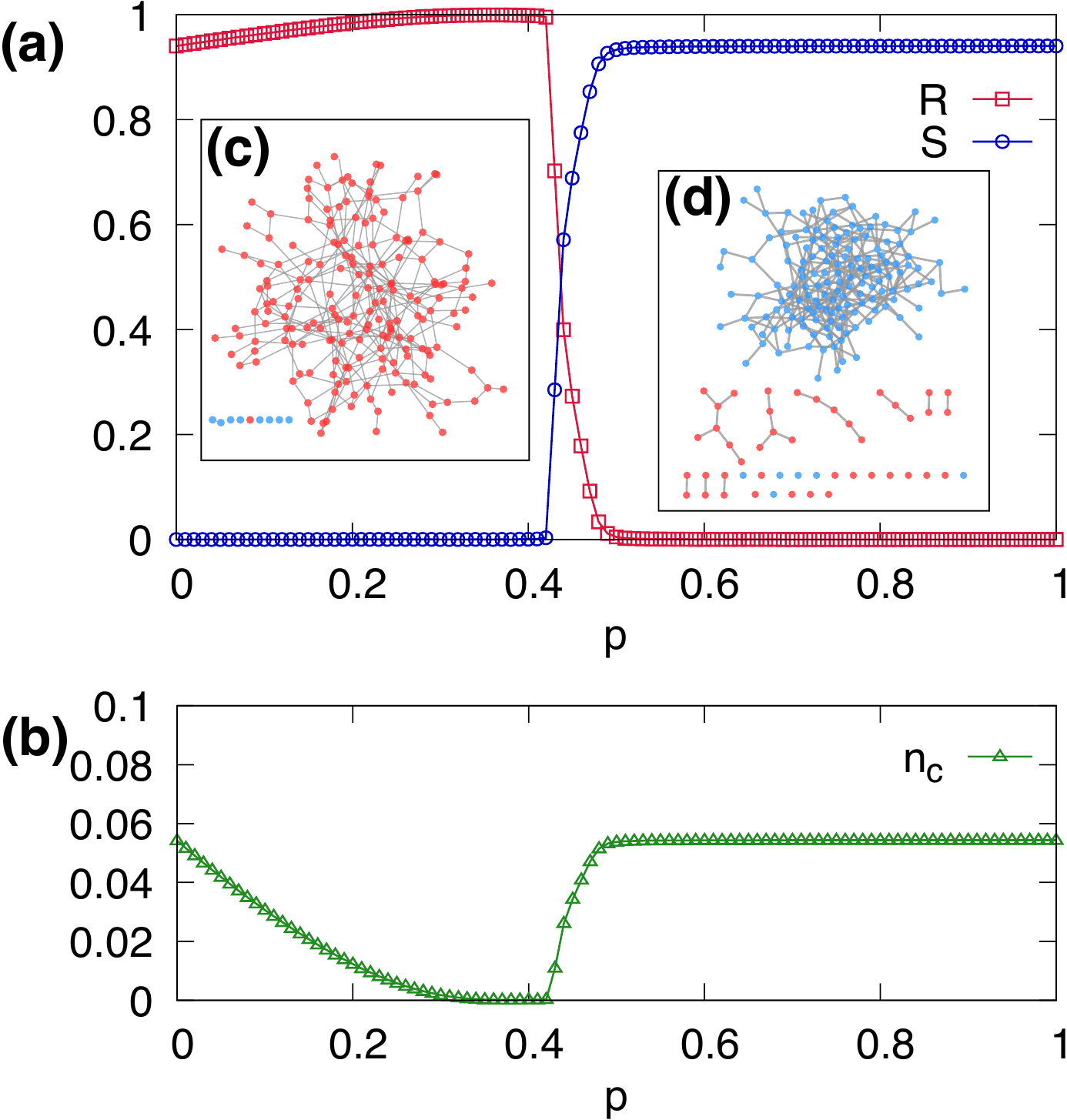}
\caption{(\textbf{a}) The final fraction $R$ of adopting nodes and the size $S$ of the largest non-adopting cluster
as a function of the network plasticity $p$. 
(\textbf{b}) The number $n_c$ of clusters to network size $N$ as a function of $p$.
The dynamics starts with $\theta=0.18$ in ER networks with $N=10^5$, $z=3$,
and an initial fraction of seeds of $R_0=2 \times 10^{-4}$.
The average values are obtained by $10^4$ independent runs with different network
realizations for each run.
Examples of network structures with $N=200$ at the steady state with (\textbf{c}) $p=0.2$ 
and (\textbf{d}) $p=0.8$. Red and blue nodes represent 
adopting and non-adopting states, respectively.
}\label{figure2}
\end{figure}

We have explored the coevolutionary threshold dynamics with link rewiring in Erd\"os--R\'enyi (ER) 
networks with $N=10^5$, $z=3$, and $\theta=0.18$. We set the initial fraction of seeds 
as $R_0=2 \times 10^{-4}$. To start with, we measured the global cascade size $R$ 
as a function of network plasticity $p$ in order to examine the effect of link rewiring. 
We also computed the size $S$ of the largest cluster composed of non-adopting nodes in order 
to inspect the network structure. The size of adopting nodes $R$ and the largest non-adopting 
cluster $S$ in a steady state is shown in Figure~\ref{figure2}a as a function of $p$. Note that the case 
of $p=0$ corresponds to the result of threshold cascading dynamics in a static network.

We found a transition between a global cascade and no cascade for a critical value  $p_c$ of the plasticity.
When $p< p_c$, most nodes are adopting,  forming a large connected cluster of adopting nodes.
Almost all nodes belong to a single cluster when $p \approx 0.4$. 
As $p$ further increases, adopting nodes are separated from the large cluster due to 
rewiring. Beyond the transition point, the cascading dynamics originating from the seed nodes 
fail to propagate throughout the entire network. As a result, many small adopting clusters 
appear and a large cluster composed of non-adopting nodes emerges.

Figure~\ref{figure2}b shows the number of clusters $n_c$ normalized to the total number of nodes $N$ 
in a steady state as a function of network plasticity $p$. For small values of $p$, $n_c$ 
decreases as $p$ increases. That is, small non-adopting clusters gradually join adopting 
clusters as $p$ increases due to rewiring. Around $p \approx 0.4$,  almost 
all nodes belong to a single adopting cluster, and therefore $n_c \approx 0$. 
As $p$ increases beyond the transition point,  adopting nodes become segregated
due to rewiring and small adopting clusters appear. Examples of network structures at 
a steady state are shown in Figure~\ref{figure2}c for $p=0.2$ and Figure~\ref{figure2}d for $p=0.8$. 
When $p=0.2$ in Figure~\ref{figure2}c, there exists a single large cluster of adopting nodes.
On the other hand, when $p=0.8$ in Figure~\ref{figure2}d, adopting nodes are segregated, resulting 
in a low $R$ value. Therefore, we find that in the ``global cascade'' phase, there is
one large adopting cluster, whereas in the ``no cascade'' phase there is a large non-adopting 
cluster and many small adopting clusters. In summary, a mechanism for the transition 
to the ``no cascade'' phase in the coevolutionary model is the  segregation of adopting nodes via link rewiring.

\subsection{Phase Diagram for Global Cascades}

\begin{figure}
\includegraphics[width=\linewidth]{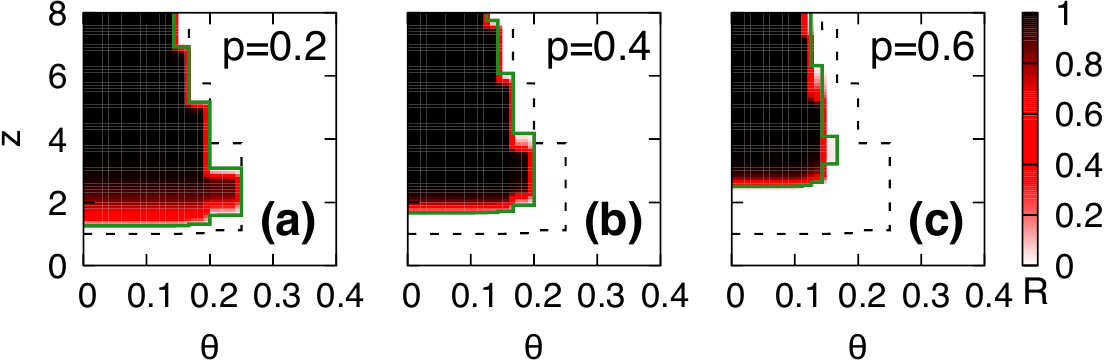}
\caption{
The final fraction $R$ of adopting nodes in ER networks with $N=10^5$ as a function of 
the average degree $z$ and threshold $\theta$, with various rewiring probabilities, i.e.,
(\textbf{a}) $p=0.2$, (\textbf{b}) $p=0.4$, and (\textbf{c})~$p=0.6$, in a steady state. 
Th dashed lines represent the transition points between the global cascade and no cascade 
phases in static networks, that is $p=0$, obtained from Equation~(\ref{pc}).
The solid lines represent the transition points with network plasticity $p$
by using mean-field approximations.
The numerical results are obtained by $10^3$ independent runs with different network
realizations for each~run.
}\label{figure3}
\end{figure}

We conducted numerical simulations to determine the fraction $R$ of the adopting nodes in the steady state
by varying the average degree $z$, the adoption threshold $\theta$, and rewiring probabilities $p=0.2, 0.4$, and $0.6$
using ER graphs with $N=10^5$ and $R_0 =2 \times 10^{-4}$ (Figure~\ref{figure3}). The dashed lines in Figure~\ref{figure3} represent 
the location of the transition lines between the ``global cascade'' and ``no cascade'' phases in static networks as 
obtained from Equation~(\ref{pc}). One of the key findings is that the domain of global cascades
shrinks  with increasing network plasticity $p$. 
Specifically, for a fixed threshold $\theta$, as $p$ increases, the first transition point 
$z_1$ of the mean degree increases, whereas the second transition point $z_2$ decreases.
The first threshold $z_1$ for the global cascades becomes delayed with increasing $p$ 
because the rewiring of links effectively segregates the adopting nodes, as described 
in the previous section. In addition, the second threshold $z_2$ decreases because the nodes 
that exceed their threshold  also become rare due to link rewiring $p$. Unlike coevolving simple 
contagion models \cite{gross2008}, the second transition $z_2$ is a peculiar feature of threshold models.

Our finding shows that link rewiring suppresses the emergence of global cascades as compared to what occurs in a static network. This is because the rewiring process removes the links 
that connect adopting and non-adopting nodes.
Consequently, the cascading dynamics become segregated  and cannot propagate further. 
Therefore, the adaptive mechanism enabled by the rewiring process effectively 
suppresses global cascades by removing active links, i.e., links 
that connect adopting and non-adopting nodes.
This mechanism allows the network to reorganize itself in response to the changes in the
state of the nodes, effectively preventing the spread of a new state.

\subsection{Non-Monotonicity in the Size of the Global Cascade}

\begin{figure}
\includegraphics[width=\linewidth]{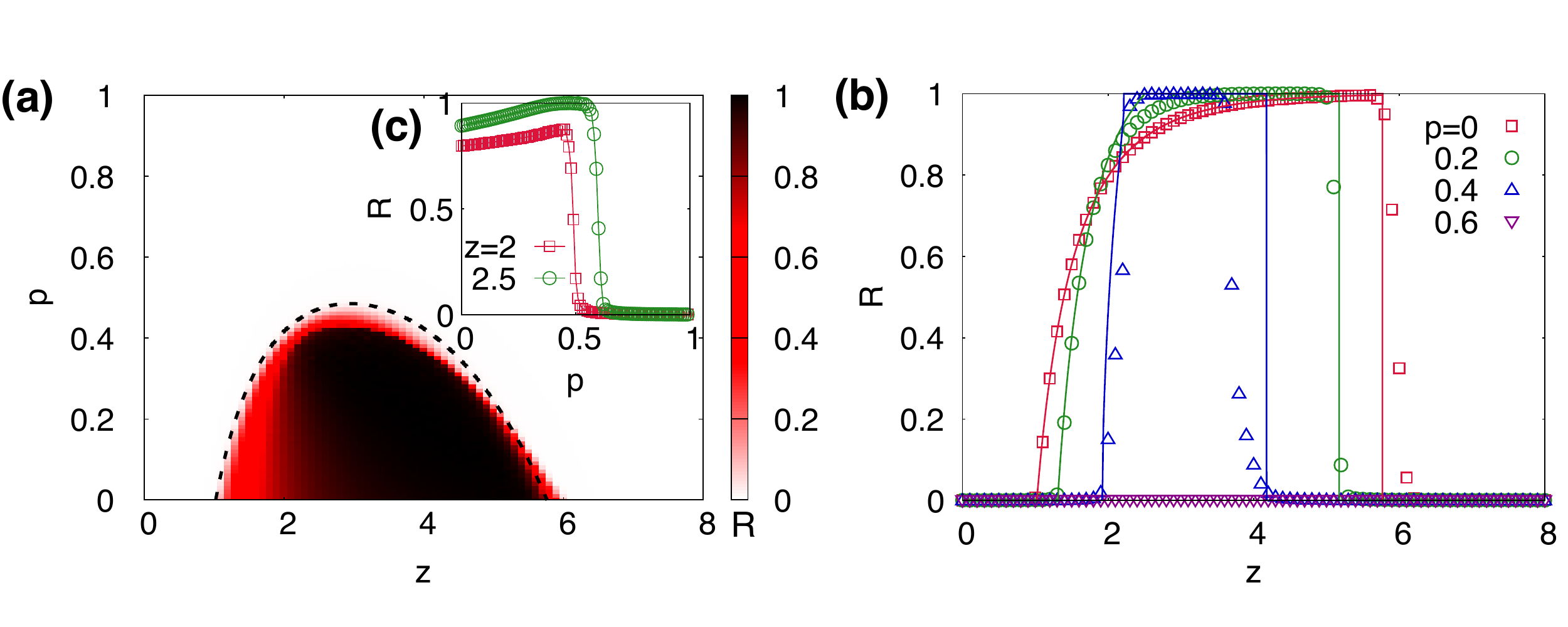}
\caption{
(\textbf{a}) The final size $R$ of cascades as a function of network plasticity $p$ 
and average degree $z$ of the ER networks with threshold $\theta=0.18$ 
and  seed fraction $R_0=2\times 10^{-4}$. 
The dashed lines represent analytical predictions obtained by Equation~(\ref{pcri}).
(\textbf{b}) The size $R$ as a function of the average degree $z$ in ER networks
for $p=0, 0.2, 0.4$, and $0.6$ with $\theta=0.18$. 
The lines represent analytical predictions based on Equations~(\ref{mf1}) and (\ref{mf2}).
(\textbf{c}) Inset shows the size $R$ with respect to the probability of link 
rewiring $p$ for $z=2$ and $2.5$ and $\theta=0.1$. 
The numerical results were obtained with $10^3$ independent runs
with different network realizations for each run.
}\label{figure4}
\end{figure}

While the area of the parameter space ($z$,$\theta$) in which global cascades occur decreases monotonically 
with the increasing network plasticity $p$, the size $R$ of the global cascades exhibits more complex patterns.
One could expect that the size $R$ also decreases monotonically with increasing $p$, 
but we found that $R$ can increase with increasing $p$ within a certain range of $p$ in
the global cascade phase. Figure~\ref{figure4}a shows the size $R$ of cascades as a function of
$p$ and $z$, with $R_0=2\times 10^{-4}$, $N=10^5$, and $\theta=0.18$ in ER networks. 
As $p$ increases, the value of $z_1$ at which the global cascades begin to occur 
is delayed. However, when the global cascade is initiated, the rate of increase 
in $R$ is greater for larger values of $p$, as shown in Figure~\ref{figure4}b. Hence, in the 
region in which $2\lesssim z \lesssim 4$, we show that the cascade size $R$ 
increases as the link rewiring probability $p$ increases. Figure~\ref{figure4}c shows
$R$ as a function of $p$ with $\theta=0.1$ and $z=2$ and $2.5$ 
 in the region where the increase in $R$ with $p$ is maximized.
In this figure, $R$ increases as the plasticity $p$ increases below the transition 
point to the ``no cascade''~phase.

The increase in $R$ with increasing $p$ can occur when separated  non-adopting clusters 
are connected to the giant cluster through new connections established during link rewiring. 
This effect can be characterized by considering the number $n_c$ of  clusters, as 
shown in Figure~\ref{figure2}b. The fraction $n_c$ of the clusters decreases 
as $p$ increases from $p=0$. This implies that increasingly more small clusters merge into the 
giant connected component of the network as $p$ increases, thereby promoting 
the larger size of cascading dynamics. For instance, when $p=0.4$, the separated  nodes 
cease to exist, indicating that initially separated  nodes have become linked to a  
cluster via link rewiring.

\subsection{Structure of Rewired Networks}

\begin{figure}
\includegraphics[width=\linewidth]{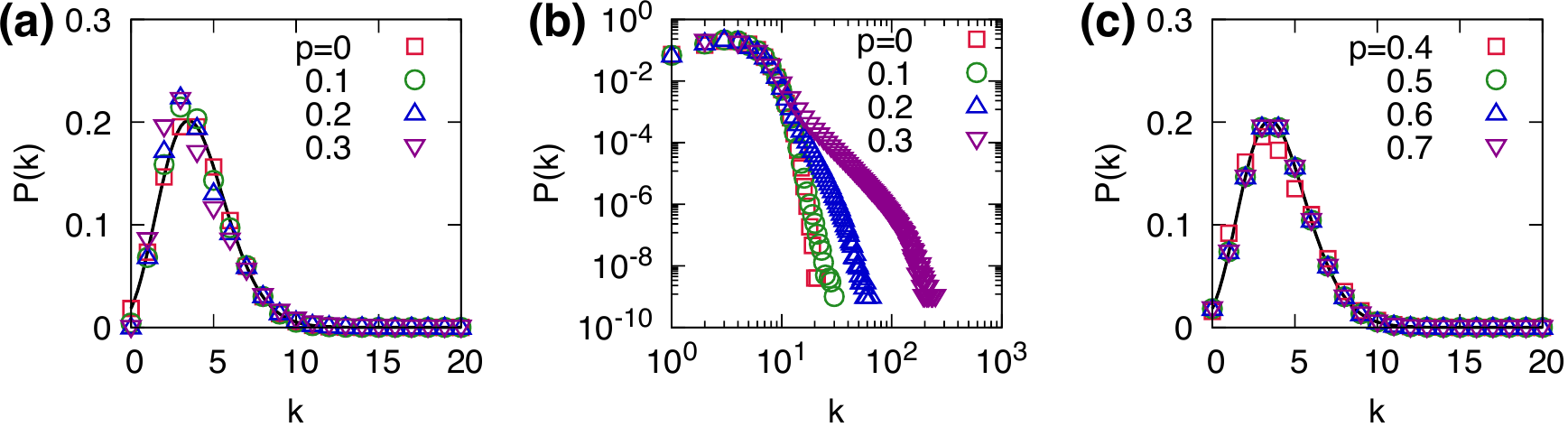}
\caption{
Degree 
 distribution $P(k)$ of the coevolving threshold model at the steady state
in (\textbf{a}) linear and (\textbf{b}) log scales for $p=0, 0.1, 0.2, 0.3$ 
(global cascade region)
and (\textbf{c}) linear scale for $p= 0.4, 0.5, 0.6, 0.7$
 (no cascades region).
The results were obtained from ER networks with $z=4$ and $N=10^5$
with $10^4$ independent runs.
The solid lines in (\textbf{a},\textbf{c}) represent the Poisson distribution with $z=4$.
}\label{figure5}
\end{figure}

We examined the network structure in the steady state. We found that the degree 
distribution broadens as $p$ increases. Figure~\ref{figure5} shows the degree distribution $P(k)$
in the steady state for various values of $p$, where the dynamics start from ER networks with $z=4$
and  $N=10^5$. As $p$ increases, the distribution deviates from a Poisson distribution as
indicated by the solid line in Figure~\ref{figure5}a. In particular, the probability of finding a 
large degree value $k$ increases with $p$, which leads to a broader degree distribution. 
This implies that as the network plasticity $p$ increases, the degree distribution 
becomes broader owing to rewiring.

In our coevolving threshold model, non-adopting nodes remove their links to adopting nodes 
and then randomly connect to a new non-adopting neighbor. Through this process, the link 
density of the non-adopting nodes continuously increases over time. 
Consequently, nodes with higher degrees gradually appear during the evolution.
Furthermore, this continuous increase in link density and the emergence of higher degree nodes 
intensify the overall connectivity of non-adopting nodes, potentially promoting the larger size of cascading dynamics.
On the other hand, in the no cascades phase, there are no adopting nodes of extensive size;
therefore, the number of active links that can be potentially rewired is limited.
Therefore, the degree distribution in this region remains approximately a Poisson 
distribution (Figure~\ref{figure5}c).

\subsection{Mean-Field Approximations}

Finally, we propose a mean-field approximation of the coevolving threshold model
on random networks that accounts for our numerical results. 
We consider the effects of link rewiring by generalizing the mean-field equations 
for the static networks. In our model, there are two main effects of link rewiring: 
removing active links between adopting and non-adopting nodes and 
increasing the density of links between non-adopting nodes as new links are created.
By implementing these two effects, we modify the self-consistency equation (Equation~(\ref{eq:qn})) for the
probability $q_n$ in a local tree-like network as
\begingroup
\makeatletter\def\f@size{9}\check@mathfonts
\def\maketag@@@#1{\hbox{\m@th\normalsize\normalfont#1}}%
\begin{align}
\label{mf1}
q_n &= (1-\tilde{p})q_0 + (1- \tilde{p})(1-q_0) \sum_{k=0}^{\infty} \frac{k Q(k,n)}{z} \\
		&\times \sum_{m=0}^{k-1} \binom{k-1}{m} q_{n-1}^m (1-q_{n-1})^{k-m-1} F(m/k, \theta), \nonumber
\end{align}\endgroup
where $Q(k,n)$ is the degree distribution of the non-adopting nodes at time step $n$
and $\tilde{p}$ is the probability that an active link will be removed before 
a non-adopting node at one end of the link is adopting.
Note that unlike the threshold model in a static network, the degree distribution is neither 
time independent nor initially given because of the rewiring processes. 
Similarly, the mean-field equation of the cascade size at step $n$ is approximately given by
\begin{align}
\label{mf2}
R_n &= R_0  +(1-R_0) \sum_{k} Q(k,n)  \\
	&\times \sum_{m=0}^{k}	\binom{k}{m} q_n^{m}(1-q_n)^{k-m} F( m/k, \theta), \nonumber
\end{align}
where $R_n$ is the fraction of adopting nodes at time step $n$.

We suggest a zero-th order estimation of the probability of  link 
removal $\tilde{p}$ and the time-dependent degree distribution $Q(k,n)$. 
If an adopting node is connected to a non-adopting node, the link between them is 
removed with probability $p$ at each time step. 
Therefore, in order to make an accurate prediction of $\tilde{p}$, it is necessary to 
know the time interval required for a non-adopting node to become adopting.
However, this interval is difficult to predict, because
the value is determined by collective interactions and not by the properties of individual links. 
To qualitatively explore the effect of link rewiring, we assume that link rewiring 
only affects one time step, leading to $\tilde{p} \approx p$. 
This assumption underestimates the actual value of $\tilde{p}$
because the active links can persist for multiple steps. However, it can qualitatively explain
the effect of active link removal as an approximation.

Subsequently, we estimated the time-dependent degree distribution $Q(k,n)$
by adding a new connection randomly among non-adopting nodes. As the time step $n$ 
increases, the average degree of non-adopting nodes also increases. In well-mixed 
populations consisting of $N$ nodes, the degree distribution of non-adopting 
nodes at  step $n$ can be approximated as follows: 
\begin{align}
\label{qq}
Q(k,n) = \binom{N_n}{k} \pi_n^k (1-\pi_n)^{N_n-k}, 
\end{align}
where $N_n$ is the number of non-adopting nodes ($N_n = (1-R_n)N$)
and $\pi_n$ is the probability that two randomly chosen non-adopting nodes 
are connected at time $n$. To estimate the probability $\pi_n$, 
we assume again that the effect of link rewiring lasts for only one time step.
Thus, the probability $\pi_n$ can be approximated as follows:
\begin{align}
\label{pi}
\pi_n = \pi_{n-1} \left[ 1 + \frac{p (1-\theta)\Delta R_{n-1}}{1-R_{n-1}} \right],
\end{align}
where $\Delta R_{n-1}$ is the change in $R$ between the steps $n-1$ and $n$. 
The estimation is based on the assumption that the number of additional links between 
the non-adopting nodes is equal to the number of links lost by the adopting nodes 
during link rewiring. The term $(1-\theta)$ represents the maximum fraction of active 
links in an adopting node that are subject to link rewiring, because at least $\theta$ 
fraction of links are already connected to adopting neighbors.

Combining Equations~(\ref{qq}) and (\ref{pi}), we can estimate the final fraction $R_{\infty}$ of 
the adopting nodes in the steady state by iteratively solving  Equations (\ref{mf1}) and (\ref{mf2}).
A comparison of the theoretical predictions for $R$
and the numerical simulation results is shown in Figure~\ref{figure4}b. The lines in Figure~\ref{figure4}b corresponding to the 
analytical results give a good  description of the main features of the numerical results. 
Our approximation accounts for the  two dynamical effects of link rewiring: one segregates
the adopting nodes by removing the active links, and the other increases the link density of 
non-adopting nodes, which could promote contagion above the first transition $z_1$.  
The quantitative discrepancies between our mean-field approximation and numerical results 
are primarily caused by the assumptions that we made to derive $\tilde{p}$, $Q(k,n)$ and 
the term $(1-\theta)$ in $\pi_n$.

In addition, the necessary condition for global cascades in the limit $R_0 \rightarrow 0$ 
can be predicted by a linear stability analysis of a trivial fixed point $q_{\infty}=0$ as follows:
\begin{align}
\label{qc}
(1- \tilde{p}) \sum_{k=1}^{\infty} \frac{ k (k-1)}{z} P(k) F(1/k,\theta) > 1.
\end{align}

\begin{figure}
\includegraphics[width=\linewidth]{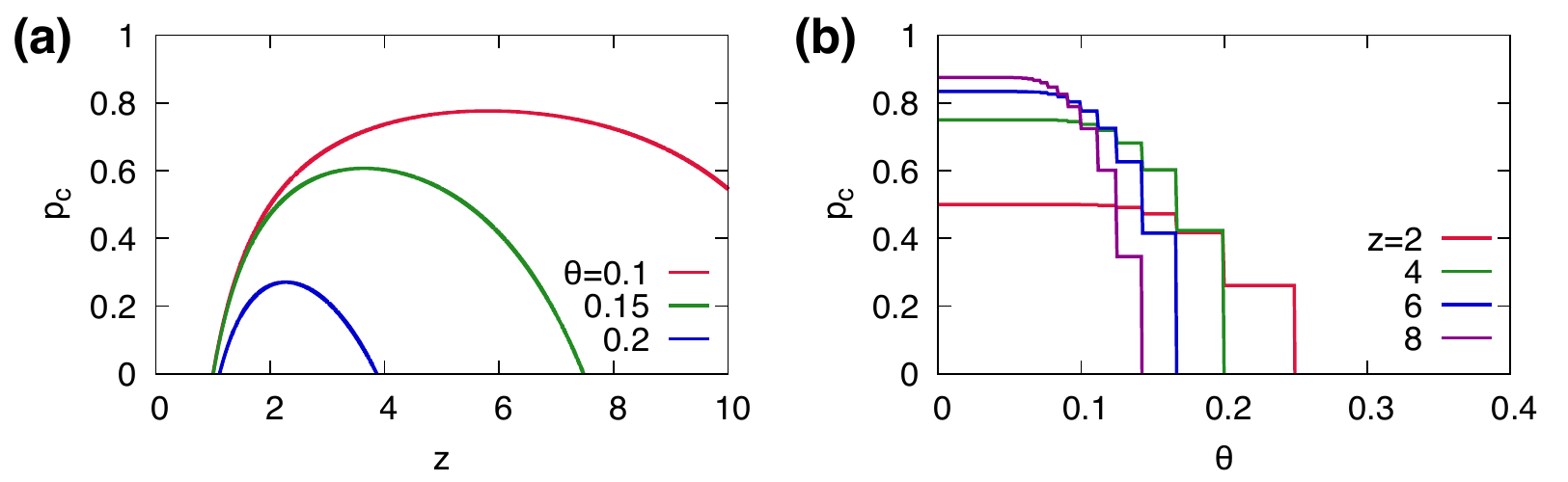}
\caption{
The critical values $p_c$ of network plasticity estimated by 
the mean-field approximation
(\textbf{a}) with respect to $z$ with fixed $\theta=0.1,0.15,0.2$ and 
(\textbf{b}) with respect to $\theta$ with fixed $z=2,4,6,8$.
}\label{figure6}
\end{figure}

In order for a global cascade to occur, an initial cascade must be triggered; hence, we estimate
the necessary condition using the degree distribution with $n=0$, $P(k)$.
The cascading condition in the coevolutionary dynamics is approximately modified by 
a factor of $(1-\tilde{p})$ from the condition in static networks. 
This implies that the cascading condition of the coevolutionary cascading model obtained 
with the mean-field approximation predicts the condition for the cascading dynamics 
that involves random link removals with a probability $\tilde{p}$.
When we approximate $\tilde{p} \approx p$, the transition point can be estimated.
The transition points between global cascade and no cascade phases predicted by the theory
are denoted by lines in Figures~\ref{figure3} and \ref{figure4}a. Overall, our mean-field approximations 
give reasonable predictions.
We can predict the critical value of network plasticity, denoted as  $p_c$, as:
\begin{align}
\label{pcri}
p_c = 1- \left[ \sum_{k=1}^{\infty} \frac{ k (k-1)}{z} P(k) F(1/k,\theta) \right]^{-1}. 
\end{align}
The critical values $p_c$ with respect to $z$ with fixed  $\theta$ 
and with respect to $\theta$ with fixed $z$ are shown in Figure~\ref{figure6}.

\section{Summary and Discussion}

We have studied the coevolutionary dynamics of network topology and social complex contagion  
using a binary-state threshold cascade model. We investigated how the mechanism for a global 
cascade is modified by the dynamics of the network topology and also the asymptotic stationary 
state of the network structure. Network dynamics, characterized by a plasticity parameter $p$, 
follow from a rewiring  of links to cut the connections between nodes in different states.  
We find that the network dynamics suppress the onset of global cascades;
there is a transition from a ``global cascade''  state to a ``no cascade'' state
as the network plasticity $p$ is increased beyond a critical value $p_c$, so that the domain 
of parameters (threshold $\theta$ and network mean degree $z$) for which global cascades occur 
shrinks  compared to the situation in a static network. 
We have found that non-adopting nodes become more densely connected during evolution, leading 
to a broader degree distribution and to a non-monotonous dependence of cascades sizes 
on plasticity $p$ within the ``global cascade'' phase. 
We have also developed a mean-field 
approximation that provides a good description of the transition lines between the ``global cascade''
and ``no cascade'' phases in the presence of link rewiring.

In previous models of coevolving voter dynamics, a generic result, different to what we find 
here, was the existence of a network fragmentation transition in two main network 
components~\cite{vazquez2008,bmin2016}. However, these studies considered binary-state models with two equivalent states, while here we consider the spreading of an initial 
minority state with threshold dynamics in which a change of state is only allowed from 
the initial majority state to the minority state. 
Additionally, unlike coevolving epidemic models with simple contagion \cite{gross2008}, once 
a node becomes adopting, it remains in that state permanently in our model. 
The consequence is that there is always a large network component and small segregated clusters, some of which are reminiscent of the shattered fragmentation transitions 
found in multilayer coevolution~\cite{diakonova2014,bmin2019}.
Overall, this study offers insights into the coevolutionary dynamics of social complex 
contagion and network evolution for an understanding of complex and evolving systems. 
In addition, this study provides a framework for studying and controlling the cascading 
phenomena in real-world systems, highlighting the importance of  the interplay between 
network dynamics and social complex contagion.

\acknowledgments{B.M. was supported by 
a National Research Foundation of Korea (NRF) grant funded 
by the Korean government (MSIT) (no.~2020R1I1A3068803). M.S.M. acknowledges financial support 
from MCIN/AEI/10.13039/501100011033 and the Fondo Europeo de Desarrollo Regional (FEDER, UE) 
through the project PACSS (RTI2018-093732-B-C21) and the Mar\'ia de Maeztu Program for units 
of Excellence in R\&D, grant CEX2021-001164-M/10.13039/501100011033.
}

\end{document}